\documentclass{article}

\usepackage{arxiv}

\usepackage[utf8]{inputenc} 
\usepackage[T1]{fontenc}    

\usepackage[colorlinks=true, allcolors=blue]{hyperref}
\newcommand{\email}[1]{\href{mailto:#1}{\tt{\nolinkurl{#1}}}}
\newcommand{\orcid}[1]{ORCID: \href{https://orcid.org/#1}{\tt{\nolinkurl{#1}}}}

\usepackage{url}            
\usepackage{booktabs}       
\usepackage{amsfonts}       
\usepackage{nicefrac}       
\usepackage{microtype}      
\usepackage{lipsum}
\usepackage{amssymb}
\usepackage{latexsym}

\usepackage{url}
\usepackage{xcolor}
\usepackage{hyperref}

\usepackage{amsmath}
\usepackage{graphicx}
\usepackage{cleveref}
\usepackage[export]{adjustbox}
\usepackage{subfig}
\usepackage{mwe}
\usepackage{graphicx}
\usepackage{pdflscape}

\title{A Modified Deep Convolutional Neural Network for Detecting COVID-19 and Pneumonia from Chest X-ray Images based on the Concatenation of Xception and ResNet50V2} 

\author{
 Mohammad Rahimzadeh \\
  School of Computer Engineering\\
  Iran University of Science and Technology, Iran\\
  \texttt{\email{mr7495@yahoo.com}} \\
  \texttt{\orcid{0000-0002-8550-8967}} \\
  \texttt{Corresponding author}
  \And
 Abolfazl Attar \\
  Department of Electrical Engineering\\
  Sharif University of Technology, Iran\\
  \texttt{\email{attar.abolfazl@ee.sharif.edu}} \\
  \texttt{\orcid{0000-0001-6727-432X}}}

\begin{document}
\maketitle

\begin{abstract}
{In this paper, we have trained several deep convolutional
networks with the introduced training techniques for classifying X-ray images into three classes:
normal, pneumonia, and COVID-19, based on two open-source datasets. Our data contains 180 X-ray images that belong to persons infected to COVID-19, so
we tried to apply methods to achieve the best possible results. In this research, we introduce some
training techniques that help the network learn better when we have an unbalanced dataset(fewer cases of COVID-19 along with more cases from other classes). We also propose a neural network that is a concatenation of Xception and ResNet50V2 networks. This
network achieved the best accuracy by utilizing multiple features extracted by two robust networks.
For evaluating our network, we have tested our network on 11302 images to report
the actual accuracy our network can achieve in real circumstances. The average accuracy of the
proposed network for detecting COVID-19 cases is 99.50\%, and the overall average accuracy for all
classes is 91.4\%.}
\end{abstract}




\section{Introduction}
\label{1}

The pervasive spread of the coronavirus around the world has quarantined many people and crippled many industries, which has had a devastating effect on human life quality. So far, the coronavirus has killed at least 1.5 million patients and at least 80,000 people \cite{worldmeters}. 
Due to the high transmissibility of coronavirus, the detection of this disease (COVID-19) plays an important role in controlling and planning to prevent it. 

{On the other hand, demographic conditions such as age and sex of individuals and many urban parameters such as temperature and humidity affect the prevalence of this disease in different parts of the world, which is more effective in spreading this disease \cite{pirouz2020investigating, pirouz2020development}.}

The lack of detective tools and the limitations in their production has slowed down the disease detection; as a result, it increases the number of patients and casualties. The incidence of other diseases and the prevalence and the number of casualties due to COVID-19 disease will decrease if it is detected quickly. 

The first step is to get the detection, recognize the symptoms of the disease, and use distinctive signs to detect the coronavirus accurately. Depending on the type of coronavirus, symptoms can range from symptoms of the common cold to fever, cough, shortness of breath, and acute respiratory problems. The patient may also have a few days of cough for no apparent reason\cite{xyz}.

Unlike SARS, coronavirus affects not only the respiratory system but also other vital organs in the body, such as the kidneys and liver \cite{uptodate}.
Symptoms of a new coronavirus leading to COVID-19 usually begin a few days after the person becomes infected, where, in some people, the symptoms may appear a little later. According to \cite{jiang2020review,who}, Respiratory problems are one of the main symptoms of COVID-19, which can be detected by the X-ray imaging of the chest.
CT scans of the chest can also show a disease that has mild symptoms, so analyzing these images can well detect the presence of the disease in suspicious people and even without symptoms at first \cite{sun2020clinical}. 
Using these data can also cover the limitations of other tools, such as the lack of diagnostic kits and the limitations of their production.
The advantage of using CT scans and X-ray images is the availability of CT scan devices and x-ray imaging systems in most hospitals and laboratories, and the ease of access to the data needed to train the network and thus detect the disease. In the absence of common symptoms such as fever, the use of CT scans and X-ray images of the chest has a relatively good ability to detect the disease \cite{an2020clinical}.

The use of specialized human beings to diagnose the disease is a common method of detecting COVID-19 in laboratories. In this method, the specialist uses the symptoms and injuries in the chest radiology image to detect COVID-19 disease from a healthy person or person that suffering from other diseases. This procedure has costs and, most importantly, long-term detection \cite{jiang2020review,song2020emerging}.

In recent years, computer vision and Deep Learning have been used to detect many different diseases and lesions in the body automatically \cite{litjens2017survey}.

Some examples are: Detection of tumor types and volume in lungs, breast, head and brain \cite{cheng2016computer,lakshmanaprabu2019optimal}, state-of-the-art bone suppression in x-rays, diabetic retinopathy classification, prostate segmentation, nodule classification \cite{litjens2017survey}, skin lesion classification, analysis of the myocardium in coronary CT angiography \cite{zreik2018deep}, sperm detection and tracking \cite{rahimzadeh2020sperm}, etc.

Given that chest CT scan {or X-ray images} analysis is one of the methods of diagnosing COVID-19, the use of computer vision and Deep Learning can play a beneficial role in diagnosing this disease. Since the disease became widespread, many researchers have used machine vision and Deep Learning methods and obtained good results.

{Due to the sensitivity of the Covid-19 diagnosis, the diagnosis's accuracy is one of the main challenges we face in our research. On the other hand, our focus is on increasing the detection efficiency due to the limited open-source data available.}

{In this article, we try to improve COVID-19 detection and decreasing wrong COVID-19 detections. This is done by combining two robust deep convolutional neural networks and optimizing the training parameters. Besides, we also propose a method for training the network when the dataset is imbalanced.}

In \cite{an2020clinical,yang2020chest}, statistical analysis of CT scans was performed by several specialists and diagnosticians, who classified the suspects into several classes for diagnosis and treatment.

Because of the superiority of computer vision and Deep Learning in the analysis of medical images, after the reliability of CT scans of the chest for COVID-19 detection, the researchers used these tools to diagnose COVID-19.
 Immediately, artificial intelligence began to detect the disease and measure the rate of infection and damage to the lungs using CT scans and the course of the disease, with promising results \cite{gozes2020rapid}.
 
 IN \cite{wang2020deep}, they have used an innovative CNN to classify and predict COVID-19 using lung CT scans.
\cite{gozes2020rapid} has used Deep Learning to detect COVID-19 and segment the lung masses caused by the coronavirus using 2D and 3D images. COVID-Net uses a lightweight residual projection-expansion-
projection-extension (PEPX) design pattern to investigate quantitative analysis and qualitative analysis \cite{wang2020covid}.In another research, pre-trained ResNet50, InceptionV3, and Inception ResNetV2 models have used with transfer learning techniques to classify Chest X-ray images normal and COVID-19 classes \cite{narin2020automatic}.
In \cite{li2020artificial}, they present COVNet for predict COVID-19 from CT scans that have segmented using U-net \cite{ronneberger2015u}.

Another research has combined the Human-In-The-Loop(HITL) strategy that involved a group of chest radiologists with deep learning-based methods to segment and measure infection in CT scans \cite{shan+2020lung}.
In \cite{xu2020deep}, they have tried to detect COVID-19 and Influenza-A-viral-pneumonia from their data; They have used classical ResNet-18 network structure to extract the features, and another Innovative CNN network uses these features by creating the location-attention oriented model to classify the data.

The paper is organized as follows: {In Section \ref{3}, we describe the proposed neural network, the dataset, and training techniques.} In Section \ref{4}, we have presented the experimental results, and then the paper is discussed in Section \ref{5}. In Section \ref{6}, we concluded our paper, and in \ref{7}, we presented the trained networks and the codes used in this research.

\section{Methodology}
\label{3}

\subsection{{Neural Networks}}
\label{30}

 {Deep convolutional neural networks have made a mutation in machine vision tasks. These layers have created advances in many field like Agriculture \cite{rahimzadeh2020introduction}, medical disease diagnosis \cite{lih2020comprehensive,wang2020celiac}, industry \cite{dekhtiar2018deep}.
The superiority of these networks comes from the robust and valuable semantic features they generate from the input data. Here the main job of the deep networks is detecting the infections and the kind of those in the X-ray images, so classifying the X-ray images into normal, pneumonia or COVID-19.
Some of the powerful and mostly used deep convolutional networks are VGG \cite{simonyan2014deep}, ResNet \cite{he2015deep}, DenseNet \cite{huang2016densely}, Inception \cite{szegedy2014going}, Xception \cite{chollet2017xception}.}

 {Xception is a deep convolutional neural network that introduced new inception layers. These inception layers are constructed from depthwise separable convolution layers, followed by a point-wise convolution layer. Xception achieved the third-best results on the ImageNet dataset \cite{deng2009imagenet} after InceptionresnetV2 \cite{szegedy2016inceptionv4} and NasNet Large \cite{zoph2017learning}.
ResNet50V2 \cite{he2016identity} is a modified version of ResNet50 that performs better than ResNet50 and ResNet101 on the Imagenet dataset. In ResNet50V2, a modification was made in the propagation formulation of the connections between blocks.
ResNet50V2 also achieves a good result on the ImageNet dataset.}

 {The pre-processed input images of our dataset are 300*300 pixels. Xception generates a 10*10*2048 feature map on its last feature extractor layer from the input image, and ResNet50V2 also produces the same size of feature map on its final layer. As both networks generate the same size of feature maps, we concatenated their features so that by using both of the inception-based layers and residual-based layers, the quality of the generated semantic features be enhanced.}

\begin{figure}[!ht]
\large
\centering
\includegraphics[width=\linewidth]{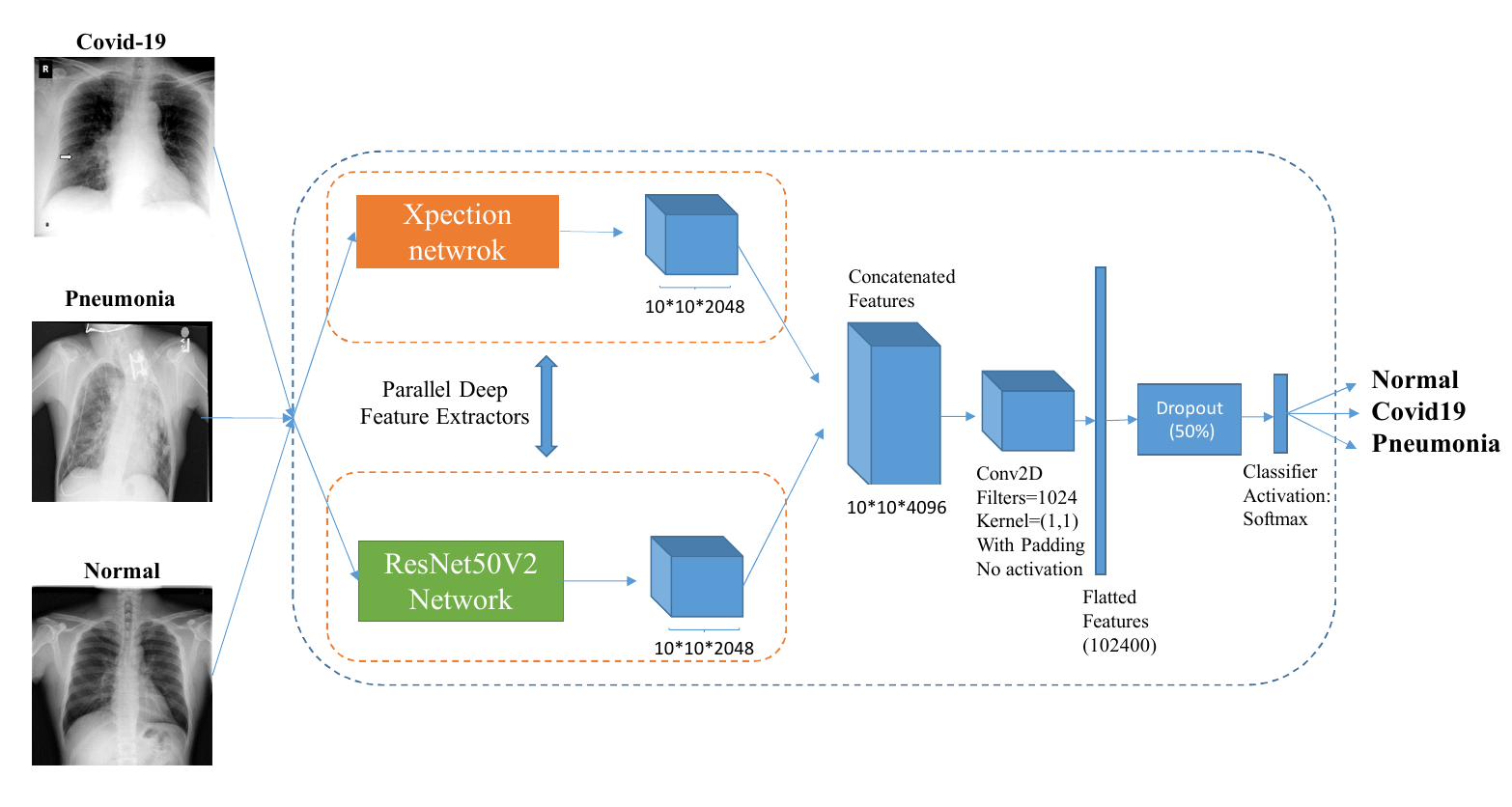}
\caption{The architecture of the concatenated network}
\label{fig1}
\end{figure}

 {The concatenated neural network is designed by concatenating the extracted features of Xception and ResNet50V2 and then connecting the concatenated features to a convolutional layer that is connected to the classifier. The kernel size of the convolutional layer that was added after the concatenated features was 1*1 with 1024 filters and no activation function. This layer was added to extract a more valuable semantic feature out of the features of a spatial point between all the channels, which each channel is a feature map. This convolutional layer helps the network learn better from the concatenated features extracted from Xception and ResNet50V2.
The architecture of the concatenated network has been depicted in figure \ref{fig1}.}

\newpage
\subsection{Dataset}
\label{31}
We have used two open-source datasets in our work. The covid chestxray
dataset is taken from this GitHub repository \newline (\url{https://github.com/ieee8023/covid-chestxray-dataset}), which has \newline been prepared by \cite{cohen2020covid}. This dataset consists of X-ray and CT scan images of patients infected to COVID-19, SARS, Streptococcus, ARDS, Pneumocystis, and other types of pneumonia from different patients.

In this dataset, we only considered the x-ray images, and in total, there were 180 images from 118 cases with COVID-19 and 42 images from 25 cases with Pneumocystis, Streptococcus, and SARS that were considered as pneumonia. The second dataset was taken from (\url{https://www.kaggle.com/c/rsna-pneumonia-detection-challenge}), which contains 6012 cases with pneumonia, and 8851 normal cases. We combined these two datasets, and the details are listed in table \ref{table1}.

\renewcommand{\arraystretch}{1.5}\

\begin{table*}[!hb]
\centering
\caption{Composition of the number of allocated images to training and validation set
in both datasets}
\label{table1}
\resizebox{\textwidth}{!}{%
\begin{tabular}{|l|l|l|l|}
\hline
Dataset                            & COVID-19 & Pneumonia & Normal \\ \hline
covid chestxray dataset            & 180      & 42        & 0      \\ \hline
rsna pneumonia detection challenge & 0        & 6012      & 8851   \\ \hline
Total                              & 180      & 6054      & 8851   \\ \hline
 {Training Set}                  & 149      & 1634      & 2000   \\ \hline
Validation Set                     & 31       & 4420      & 6851   \\ \hline
\end{tabular}}
\end{table*}

 {As stated, we only had 180 cases infected to  COVID-19, which is almost a few data for a class compared to other classes. If we had combined lots of images from normal or pneumonia classes with few COVID-19 images for training, the network would become able to detect pneumonia and normal classes very well, but not the COVID-19 class because of the unbalanced dataset. In that case, although the network can not identify COVID-19 properly, as there are many more images of pneumonia and normal classes than the COVID-19 class, the general accuracy would become very high, not the COVID-19 detection accuracy. This condition is not our goal because the main purpose here is to achieve good results in detecting COVID-19 cases and not identifying wrong COVID-19 cases.}

 {The best way to solve this problem is to make the dataset balanced and give the network almost equal data of each class when training, so the network learn to identify all the classes. Here because we do not access more open-source datasets of COVID-19 to increase this class data, we chose the number of pneumonia and normal classes almost equal to the COVID-19 number of images.}

  {We decided to train the networks for 8 consecutive phases. In each phase, we selected 250 cases of normal class and 234 cases of pneumonia class along with the 149 COVID-19 cases. In total, we had 633 cases for each training phase. All the COVID-19 images and 34 of the pneumonia images were common between each training phase and 250 normal cases, and 200 pneumonia cases were unique in each training phase. The common 149 COVID-19 and 34 pneumonia cases between all the training phases were from the covid chestxray
dataset \cite{cohen2020covid}, and the rest of the data were from the other dataset. Based on this categorizing, our training set includes 8 phases and 3783 images.}

 {By doing so, the network sees an almost equal number of images for each class, so it helps to improve the COVID-19 detection along detecting pneumonia and normal cases. But as we had more pneumonia and normal cases, we showed the network different pneumonia and normal cases with COVID-19 cases in each phase. Implementing this method results in two advantages. One is that the network learns COVID-19 class features better along with the other classes; second, the normal and pneumonia classes' detection improves very much. Better detecting pneumonia and normal cases mean not detecting wrong COViD-19 cases, which is one of our goals. In another meaning running this method helps the network better identify COVID-19 and not detect faulty COVID-19 cases.}

\begin{figure}[!ht]
\centering
\subfloat[][normal persons]{\label{main:a}\includegraphics[width=0.333\linewidth]{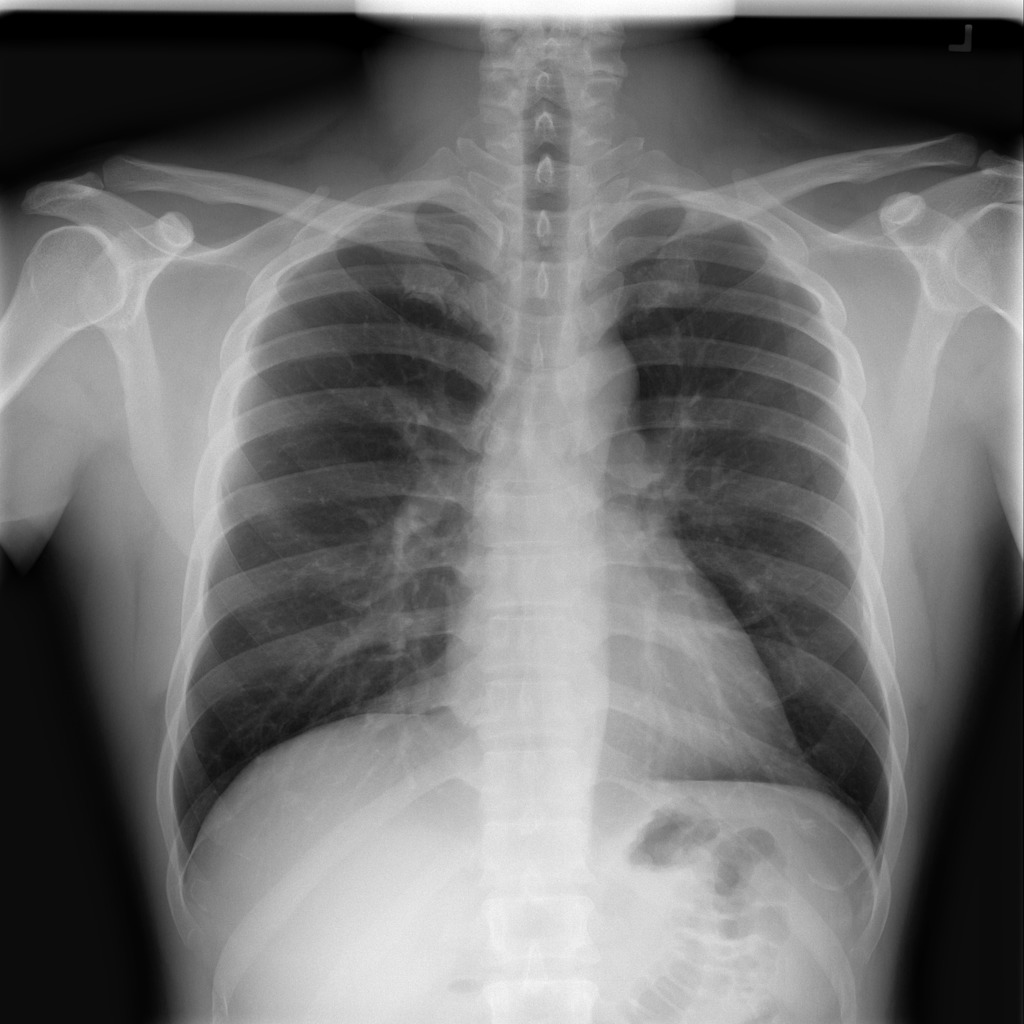}
\includegraphics[width=0.333\linewidth]{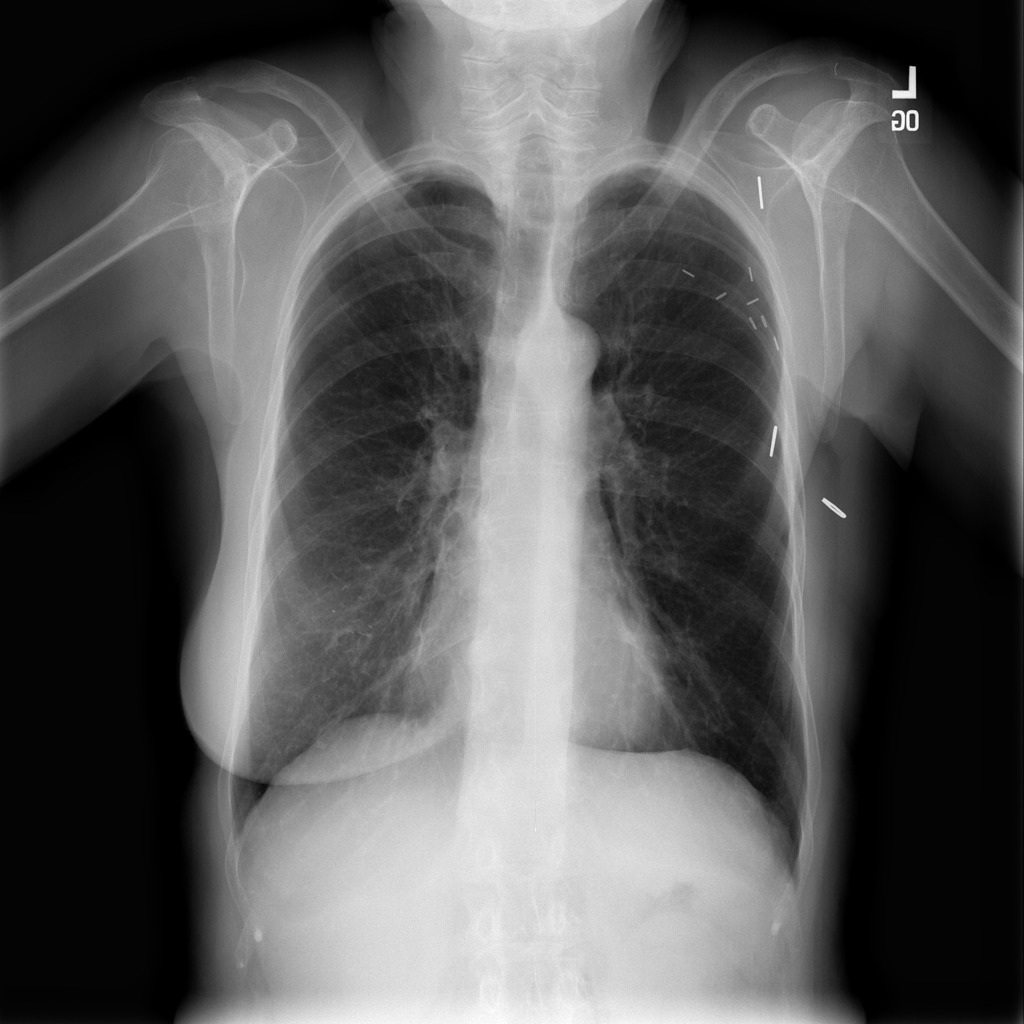}}

\subfloat[][Patients with COVID-19]{\label{main:b}\includegraphics[width=0.333\linewidth]{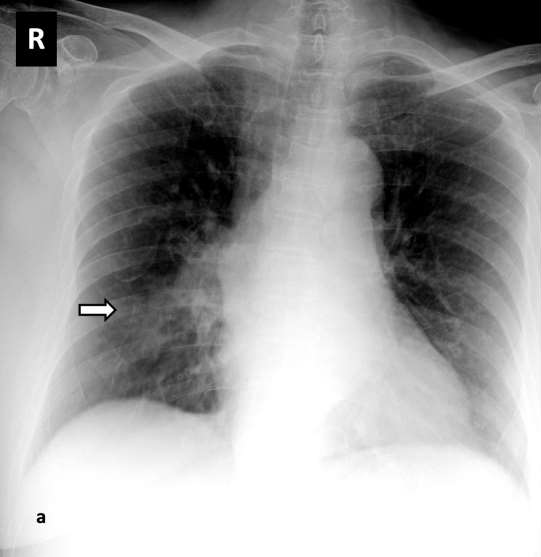}
\includegraphics[width=0.333\linewidth,height=5.65cm]{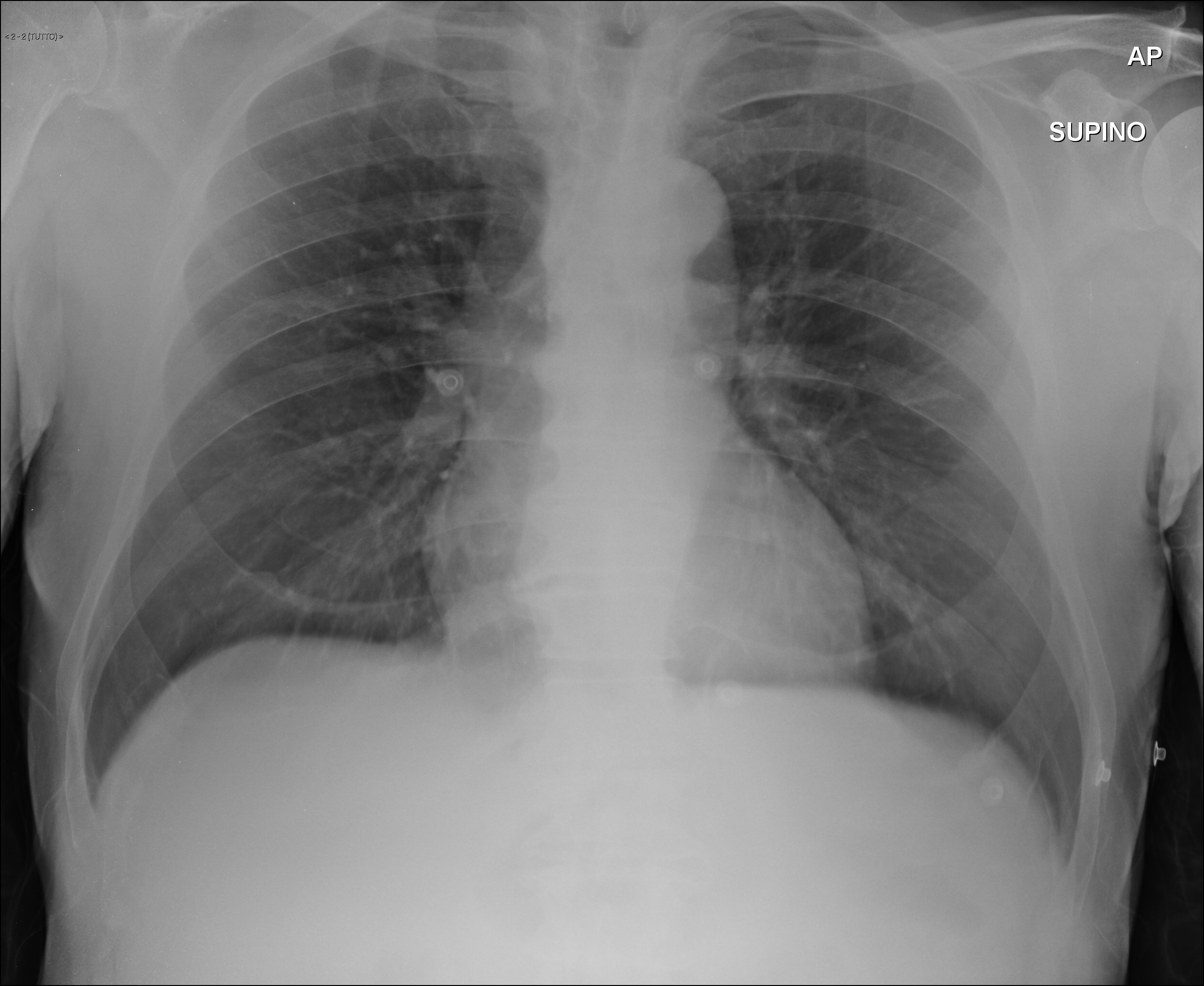}}

\subfloat[][Patients with pneumonia]{\label{main:c}\includegraphics[width=0.333\linewidth]{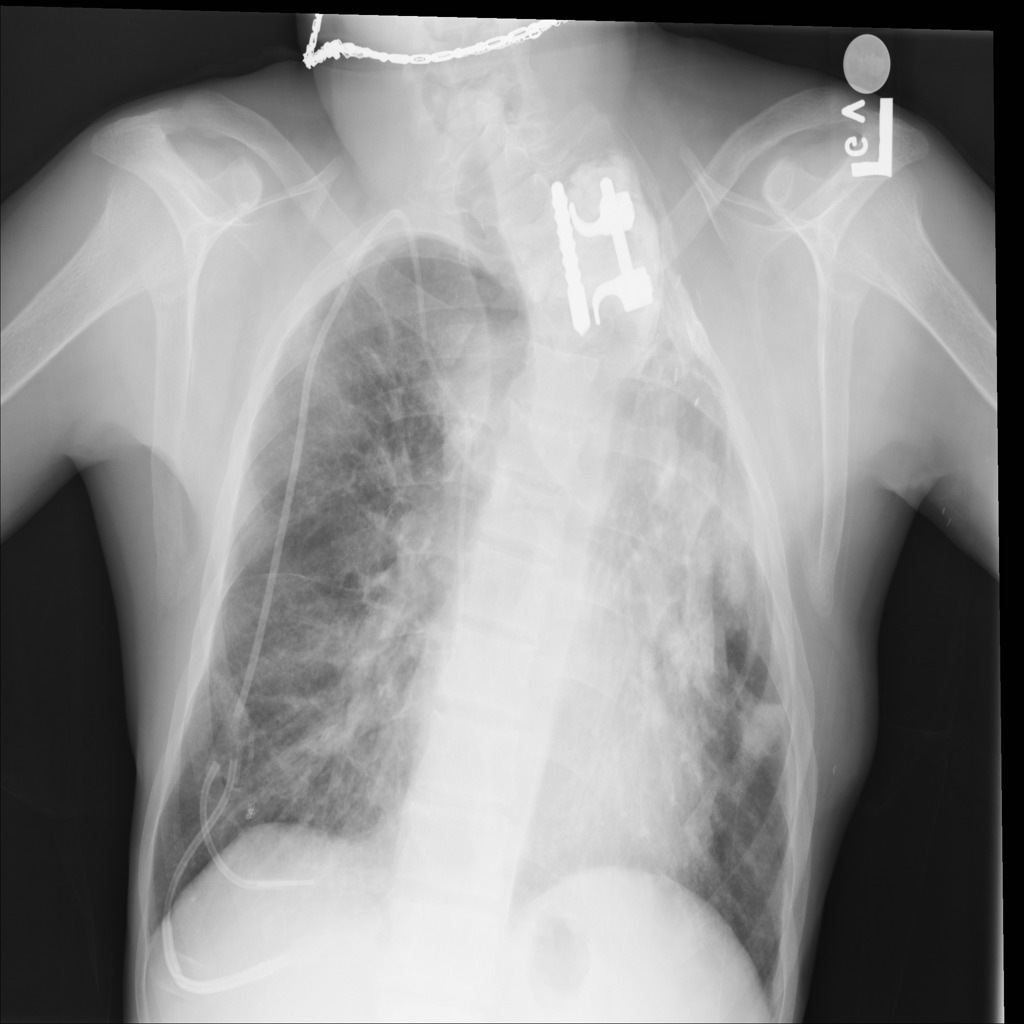}\includegraphics[width=0.333\linewidth]{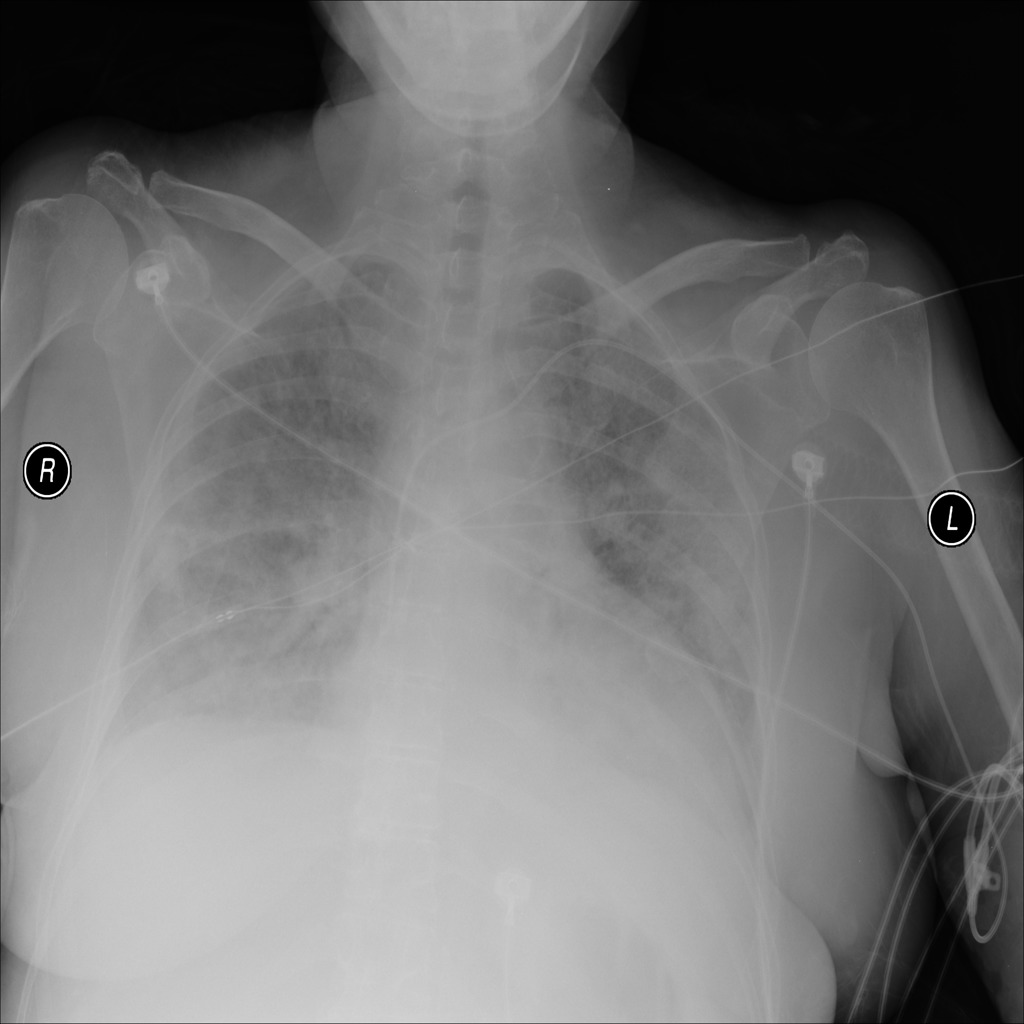}}
\caption{Examples of the images in our dataset}
\label{fig7}
\end{figure}

\begin{figure}
\centering
\includegraphics[scale=0.95]{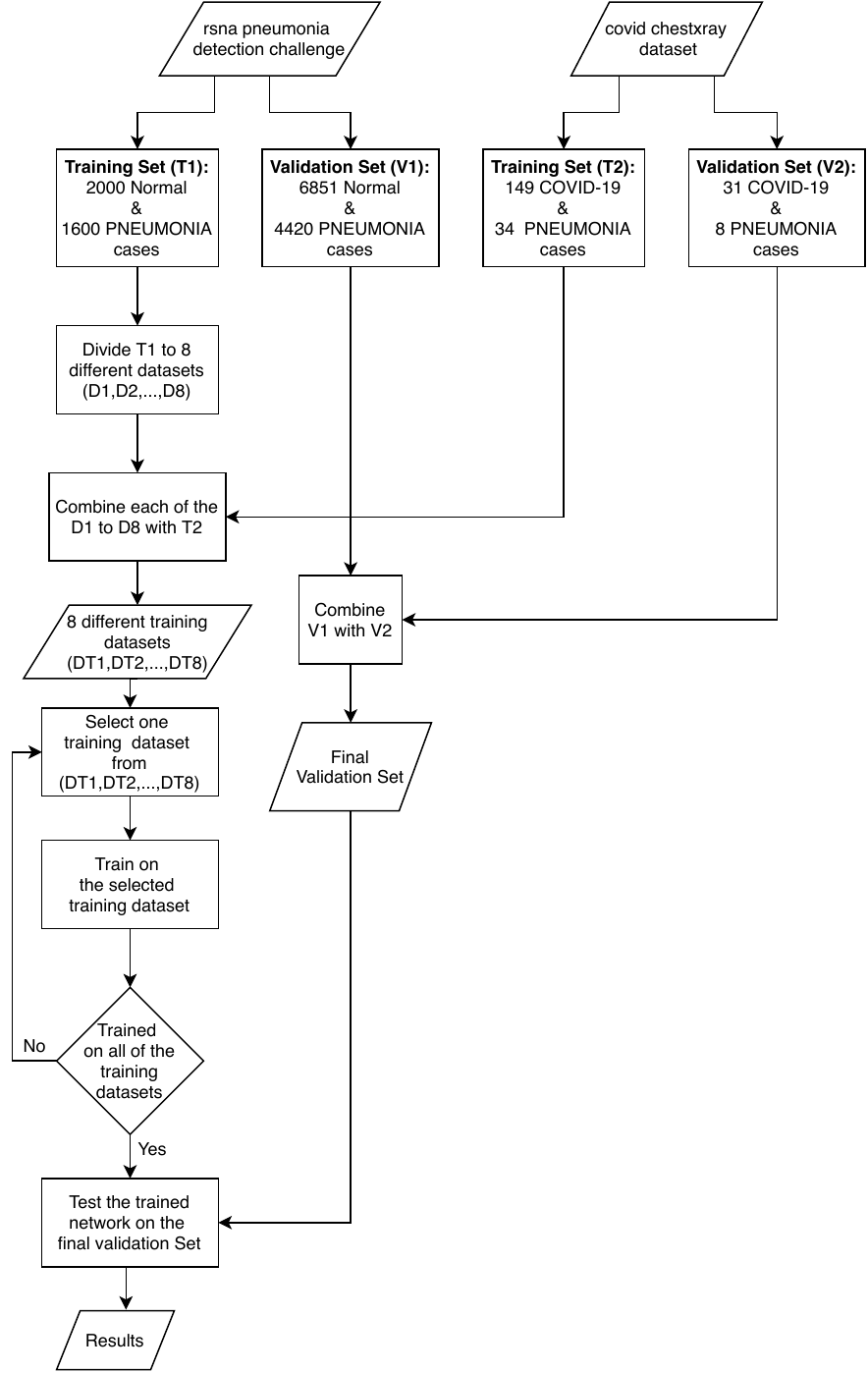}
\caption{ {The flowchart of the proposed method for training set preperation}}
\label{flow}
\end{figure}

 {This method can be used for all the circumstances that there is a highly unbalanced dataset.
We presented our way of allocating the images of the datasets into eight different phases in flowchart \ref{flow}.
Some of the images of our dataset are shown in figure \ref{fig7}.}

\subsection{Training phase}
\label{32}
 {We described in the dataset subsection \ref{31} that we allocated 8 phases for training. 
For reporting more reliable results, we chose five folds for training, which in every fold the training set was made of 8 phases as it is mentioned.}

We have trained ResNet50V2 \cite{he2016identity}, Xception \cite{chollet2017xception}, and a concatenation of Xception and ResNet50V2 neural networks based on the explained method.  This concatenated Neural Network has shown higher accuracy comparing to the others. As we have tested several networks in our project, the Xception \cite{chollet2017xception} and ResNet50V2 \cite{he2016identity} networks work almost better than other ones in extracting deep features. By concatenating the output features of both networks, we helped the network learn to classify the input image from both feature vectors, which resulted in better accuracy.
The training parameters have been described in table \ref{table2}.

 {Based on table \ref{table2} we trained the networks using Categorical cross-entropy loss function and Nadam optimizer. The learning rate was set to 1e-4. We trained the network for 100 epochs in each training phase and, because of having 8 training phases, the models were trained for 800 epochs.
For the Xception and ResNet50V2, we selected the batch size equal to 30. But as the concatenated network had more parameters than Xception and ResNet50V2, we set the batch size equal to 20. Data augmentation methods were also implemented to increase training efficiency and prevent the model from overfitting.}

We implemented the neural networks with Keras \cite{chollet2015keras} library on a Tesla P100 GPU and 25GB RAM that were provided by \href{https://colab.research.google.com/}{Google Colaboratory Notebooks}.

\begin{table*}[!ht]
\caption{In this table, we have listed the parameters and functions we used in the training procedure.}
\centering
\begin{tabular}{|l|l|l|l|}
\hline
Training Parameters                                                      & Xception                                                          & ResNet50V2                                                        & \begin{tabular}[c]{@{}l@{}}Concatenated\\ Network\end{tabular}    \\ \hline
Learning Rate                                                            & 1e-4                                                              & 1e-4                                                              & 1e-4                                                              \\ \hline
Batch Size                                                               & 30                                                                & 30                                                                & 20                                                                \\ \hline
Optimizer                                                                & Nadam                                                             & Nadam                                                             & Nadam                                                             \\ \hline
Loss Function                                                            & \begin{tabular}[c]{@{}l@{}}Categorical\\ Crossentopy\end{tabular} & \begin{tabular}[c]{@{}l@{}}Categorical\\ Crossentopy\end{tabular} & \begin{tabular}[c]{@{}l@{}}Categorical\\ Crossentopy\end{tabular} \\ \hline
\begin{tabular}[c]{@{}l@{}}Epochs per each\\ Training Phase\end{tabular} & 100                                                               & 100                                                               & 100                                                               \\ \hline
\begin{tabular}[c]{@{}l@{}}Horizontal/Vertical\\ flipping\end{tabular}   & Yes                                                               & Yes                                                               & Yes                                                               \\ \hline
Zoom Range                                                               & 5\%                                                               & 5\%                                                               & 5\%                                                               \\ \hline
Rotation Range                                                           & 0 - 360 degree                                                    & 0 - 360 degree                                                    & 0 - 360 degree                                                    \\ \hline
\begin{tabular}[c]{@{}l@{}}Width / Height\\ Shifting\end{tabular}        & 5\%                                                               & 5\%                                                               & 5\%                                                               \\ \hline
Shift Range                                                              & 5\%                                                               & 5\%                                                               & 5\%                                                               \\ \hline
Re-scaling                                                               & 1/255                                                             & 1/255                                                             & 1/255                                                             \\ \hline
\end{tabular}
\label{table2}
\end{table*}


\section{Results}
\label{4}

We validated our networks on 31 cases of COVID-19, 4420 cases of pneumonia, and 6851 normal cases. The reason our training data was less than validation data is that we had a few cases of COVID-19 among lots of normal and with pneumonia cases. Therefore,  {we could not use lots of images of the two other classes with COVID-19 fewer cases for training because it would have made the network not to learn COVID-19 features. 
To solve this issue, we selected 3783 images for training in 8 different phases. We evaluated our network on the rest of the data so that our trained network's ultimate performance be clear.}
It must be noticed that exceptionally, in the fold3, we had 30 cases of COVID-19 for validation, and 150 other cases were allocated for training.

It is noteworthy that we used transfer learning in training precess. For all the networks, we used the pre-trained ImageNet weights \cite{deng2009imagenet} at the beginning of the training and then resumed training based on the explained conditions on our dataset.
 {We also used the accuracy metric for monitoring the network results on the validation set after each epoch to find the best and most converged version of the trained network.}

\begin{figure*}[!hp]

\centering
\subfloat[Concatenated network-Fold1]{\label{main:aaa}\includegraphics[width=0.5\linewidth]{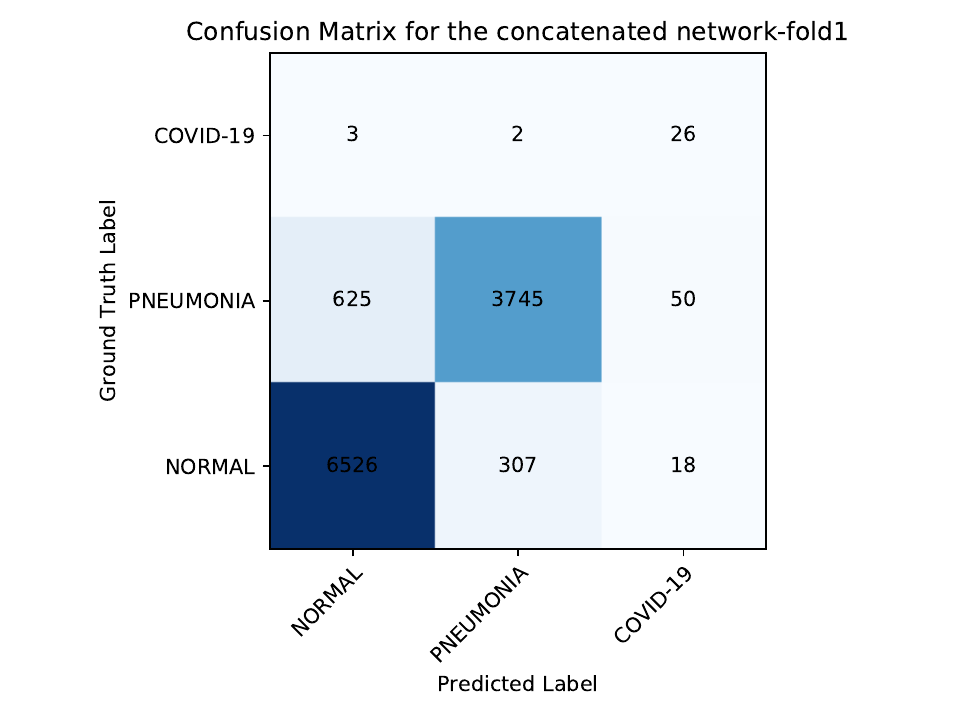}}
\subfloat[Xception-Fold1]{\label{main:bbb}\includegraphics[width=0.5\linewidth]{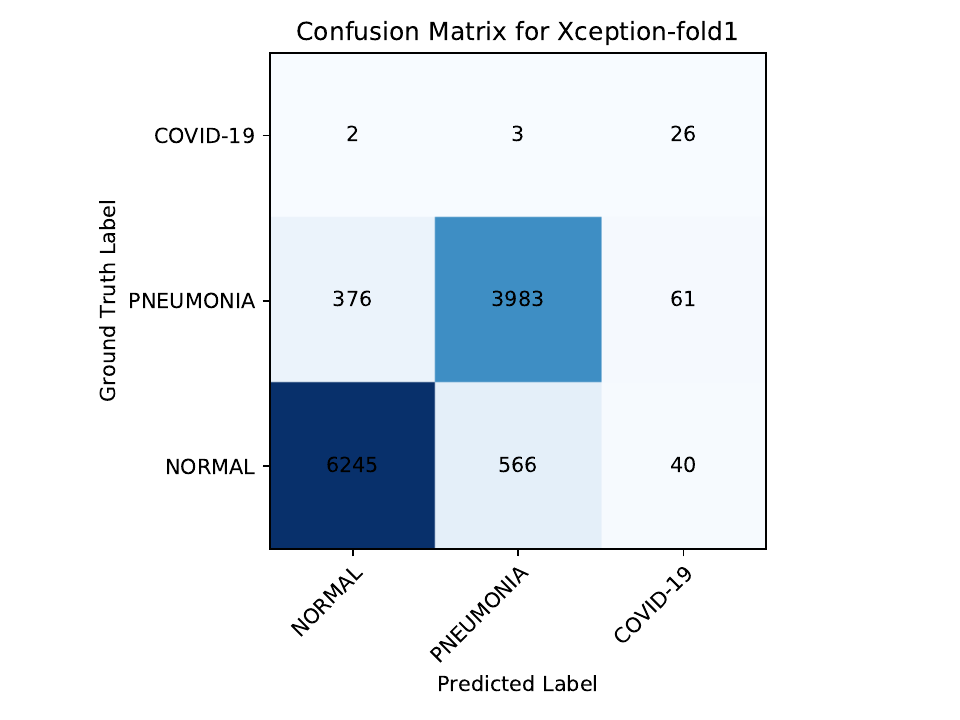}}
\newline
\subfloat[ResNet50V2-Fold1]{\label{main:ccc}\includegraphics[width=0.5\linewidth]{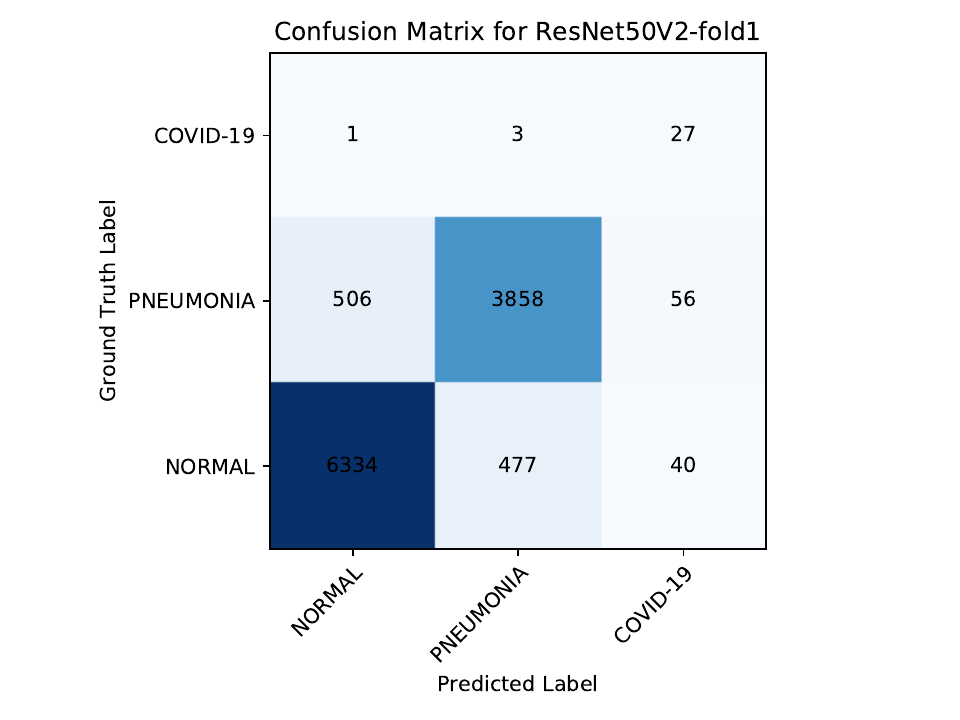}}
\subfloat[Concatenated network-Fold3]{\label{main:eee}\includegraphics[width=0.5\linewidth]{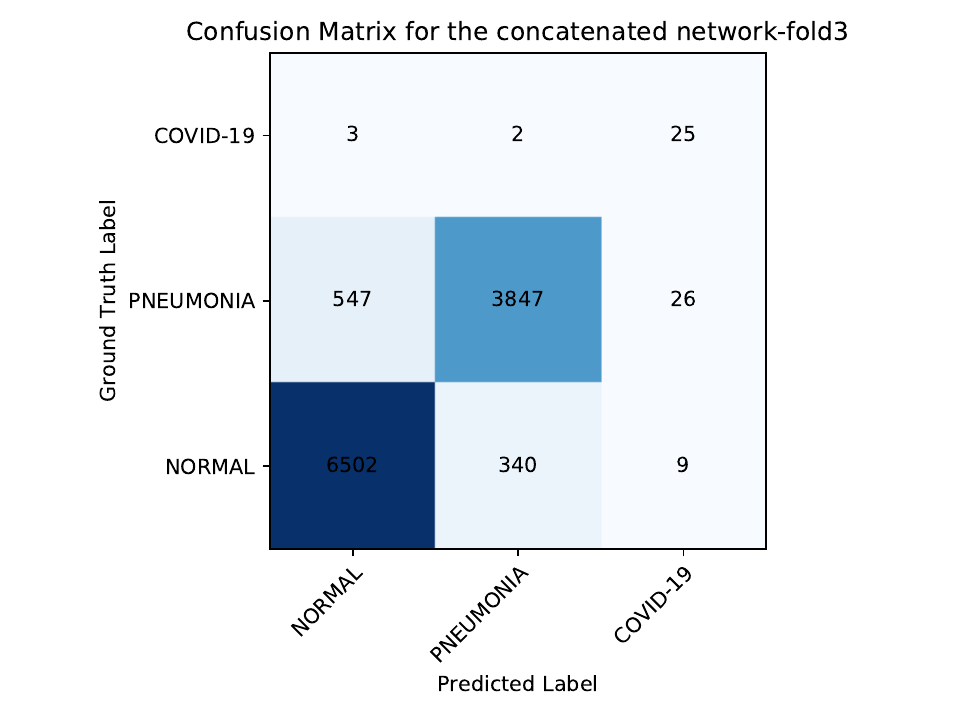}}
\newline
\subfloat[Xception-Fold3]{\label{main:fff}\includegraphics[width=0.5\linewidth]{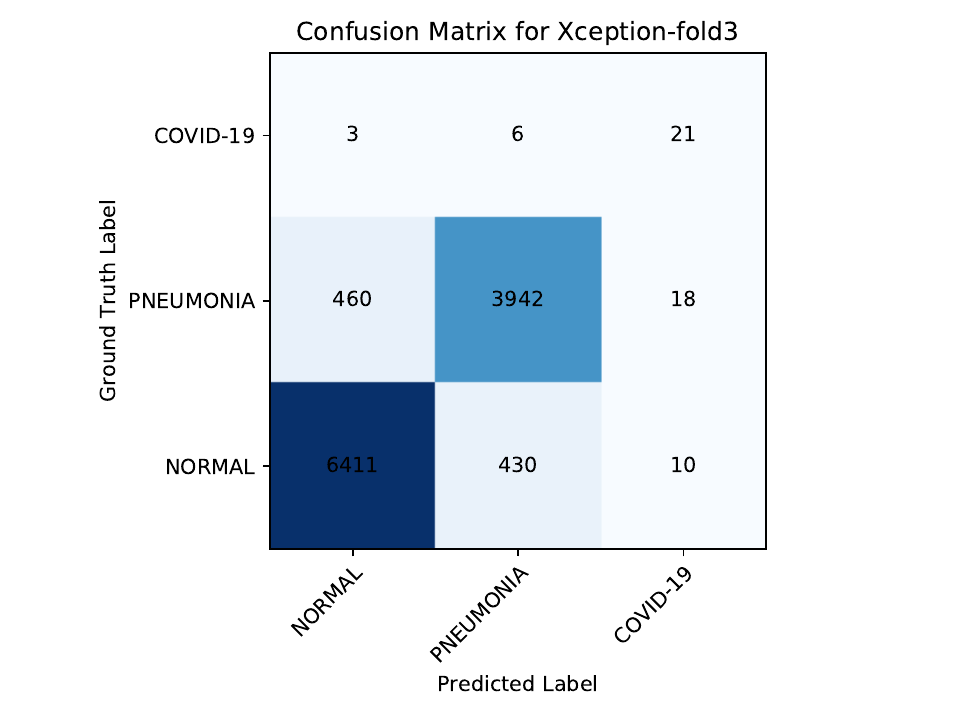}}
\subfloat[ResNet50V2-Fold3]{\label{main:ggg}\includegraphics[width=0.5\linewidth]{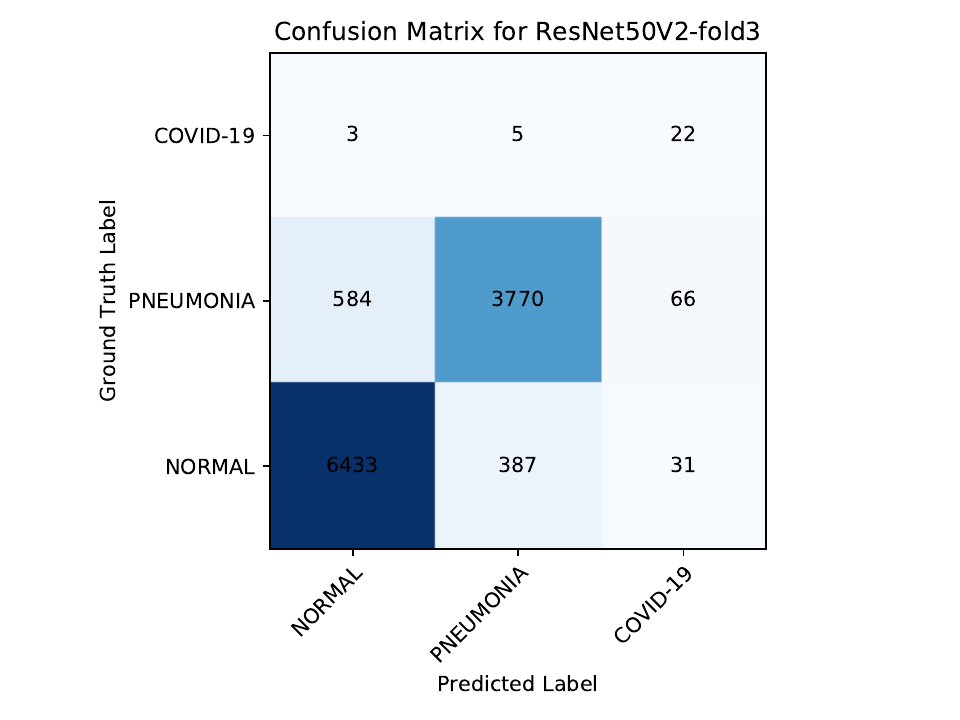}}

\caption{This figure shows the confusion matrix of the network for fold 1 and 3}
\label{fig5}
\end{figure*}

The evaluation results of the neural networks are presented in figure \ref{fig5} that shows the confusion matrices of each network for fold one and three.  {Table \ref{table3} and table \ref{table4}} show the details of our results. 
We reported the four different metrics for evaluating our network for each of the three class as follows:

\begin{equation}
 {Accuracy\ (for\ each\ class)}=\frac{TP+TN}{TP+FP+TN+FN}\label{eq:2}
\end{equation}

\begin{equation}
 {Specificity}=\frac{TN}{TN+FP}\label{eq:3}
\end{equation}

\begin{equation}
 {sensitivity}=\frac{TP}{TP+FN}\label{eq:4}
\end{equation}

\begin{equation}
 {Precision}=\frac{TP}{TP+FP}\label{eq:5}
\end{equation}

 {We also reported the overall accuracy metric, defined as:}

\begin{equation}
Accuracy\ (for\ all\ the\ classes)=\frac{Number\ of\ Correct\ Classified\  Images}{Number\ of\ All\ Images}\label{eq:1}
\end{equation}

In these equations, \(TP\) (True Positive) is the number of correctly classified images of a class, \(FP\) (False Positive) is the number of the wrong classified images of a class, \(FN\) (False Negative) is the number of images of a class that have been detected as another class, and \(TN\) (True Negative) is the number of images that do not belong to a class and did not be classified as that class.

\renewcommand{\arraystretch}{2}

\begin{table*}[!hb]
\caption{This table reports the number of true and false positives and false negatives for each class}
\label{table3}
\resizebox{\textwidth}{!}{%
\begin{tabular}{|l|l|l|l|l|l|l|l|l|l|l|}
\hline
Fold & Network      & \begin{tabular}[c]{@{}l@{}}COVID-19\\ Correct\\ detected\end{tabular} & \begin{tabular}[c]{@{}l@{}}COVID-19\\ Not\\ detected\end{tabular} & \begin{tabular}[c]{@{}l@{}}COVID-19\\ Wrong\\ detected\end{tabular} & \begin{tabular}[c]{@{}l@{}}PNEUMONIA\\ Correct\\ detected\end{tabular} & \begin{tabular}[c]{@{}l@{}}PNEUMONIA\\ Not\\ detected\end{tabular} & \begin{tabular}[c]{@{}l@{}}PNEUMONIA\\ Wrong\\ detected\end{tabular} & \begin{tabular}[c]{@{}l@{}}NORMAL\\ Correct\\ detected\end{tabular} & \begin{tabular}[c]{@{}l@{}}NORMAL\\ Not\\ detected\end{tabular} & \begin{tabular}[c]{@{}l@{}}NORMAL\\ Wrong\\ detected\end{tabular} \\ \hline
     & Xception     & 26                                                                    & 5                                                                 & 101                                                                 & 3983                                                                   & 437                                                                & 569                                                                  & 6245                                                                & 606                                                             & 378                                                               \\ \cline{2-11} 
1    & ResNet50V2   & 27                                                                    & 4                                                                 & 96                                                                  & 3858                                                                   & 562                                                                & 480                                                                  & 6334                                                                & 517                                                             & 507                                                               \\ \cline{2-11} 
     & Concatenated & 26                                                                    & 5                                                                 & 68                                                                  & 3745                                                                   & 675                                                                & 309                                                                  & 6526                                                                & 325                                                             & 628                                                               \\ \hline
     & Xception     & 23                                                                    & 8                                                                 & 42                                                                  & 3874                                                                   & 546                                                                & 409                                                                  & 6426                                                                & 425                                                             & 528                                                               \\ \cline{2-11} 
2    & ResNet50V2   & 22                                                                    & 9                                                                 & 67                                                                  & 3659                                                                   & 761                                                                & 501                                                                  & 6340                                                                & 511                                                             & 713                                                               \\ \cline{2-11} 
     & Concatenated & 23                                                                    & 8                                                                 & 27                                                                  & 3913                                                                   & 507                                                                & 434                                                                  & 6413                                                                & 438                                                             & 492                                                               \\ \hline
     & Xception     & 21                                                                    & 9                                                                 & 28                                                                  & 3942                                                                   & 478                                                                & 436                                                                  & 6411                                                                & 440                                                             & 463                                                               \\ \cline{2-11} 
3    & ResNet50V2   & 22                                                                    & 8                                                                 & 97                                                                  & 3770                                                                   & 650                                                                & 392                                                                  & 6433                                                                & 418                                                             & 587                                                               \\ \cline{2-11} 
     & Concatenated & 25                                                                    & 5                                                                 & 35                                                                  & 3847                                                                   & 573                                                                & 342                                                                  & 6502                                                                & 349                                                             & 550                                                               \\ \hline
     & Xception     & 22                                                                    & 9                                                                 & 42                                                                  & 3818                                                                   & 602                                                                & 433                                                                  & 6411                                                                & 440                                                             & 576                                                               \\ \cline{2-11} 
4    & ResNet50V2   & 22                                                                    & 9                                                                 & 78                                                                  & 4015                                                                   & 405                                                                & 758                                                                  & 6065                                                                & 786                                                             & 364                                                               \\ \cline{2-11} 
     & Concatenated & 26                                                                    & 5                                                                 & 77                                                                  & 3860                                                                   & 560                                                                & 480                                                                  & 6340                                                                & 511                                                             & 519                                                               \\ \hline
     & Xception     & 21                                                                    & 10                                                                & 41                                                                  & 4041                                                                   & 379                                                                & 502                                                                  & 6335                                                                & 516                                                             & 362                                                               \\ \cline{2-11} 
5    & ResNet50V2   & 21                                                                    & 10                                                                & 42                                                                  & 3604                                                                   & 816                                                                & 284                                                                  & 6549                                                                & 302                                                             & 802                                                               \\ \cline{2-11} 
     & Concatenated & 24                                                                    & 7                                                                 & 43                                                                  & 3941                                                                   & 479                                                                & 390                                                                  & 6442                                                                & 409                                                             & 462                                                               \\ \hline
\end{tabular}}
\end{table*}

\begin{table*}[!ht]
\caption{Some of the evaluation metrics have been reported in this table.}
\label{table4}
\resizebox{\textwidth}{!}{%
\begin{tabular}{|l|l|l|l|l|l|l|l|l|l|l|l|l|l|l|}
\hline
Fold    & Network      & Accuracy & \begin{tabular}[c]{@{}l@{}}COVID-19\\  {sensitivity}\end{tabular} & \begin{tabular}[c]{@{}l@{}}PNEUMONIA\\  {sensitivity}\end{tabular} & \begin{tabular}[c]{@{}l@{}}NORMAL\\  {sensitivity}\end{tabular} & \begin{tabular}[c]{@{}l@{}}COVID-19\\ Specificity\end{tabular} & \begin{tabular}[c]{@{}l@{}}PNEUMONIA\\ Specificity\end{tabular} & \begin{tabular}[c]{@{}l@{}}NORMAL\\ Specificity\end{tabular} & \begin{tabular}[c]{@{}l@{}}COVID-19\\ Accuracy\end{tabular} & \begin{tabular}[c]{@{}l@{}}PNEUMONIA\\ Accuracy\end{tabular} & \begin{tabular}[c]{@{}l@{}}NORMAL\\ Accuracy\end{tabular} & \begin{tabular}[c]{@{}l@{}}COVID-19\\ Precision\end{tabular} & \begin{tabular}[c]{@{}l@{}}PNEUMONIA\\ Precision\end{tabular} & \begin{tabular}[c]{@{}l@{}}NORMAL\\ Precision\end{tabular} \\ \hline
        & Xception     & 90.72    & 83.87                                                     & 90.11                                                      & 91.15                                                   & 99.1                                                           & 91.73                                                           & 91.51                                                        & 99.06                                                       & 91.10                                                        & 91.29                                                     & 20.47                                                        & 87.50                                                         & 94.29                                                      \\ \cline{2-15} 
1       & ResNet50V2   & 90.41    & 87.09                                                     & 87.28                                                      & 92.45                                                   & 99.15                                                          & 93.03                                                           & 88.61                                                        & 99.12                                                       & 90.78                                                        & 90.94                                                     & 21.95                                                        & 88.93                                                         & 92.58                                                      \\ \cline{2-15} 
        & Concatenated & 91.10    & 83.87                                                     & 84.72                                                      & 95.25                                                   & 99.4                                                           & 95.51                                                           & 85.89                                                        & 99.35                                                       & 91.29                                                        & 91.57                                                     & 27.65                                                        & 92.37                                                         & 91.22                                                      \\ \hline
        & Xception     & 91.33    & 74.19                                                     & 87.64                                                      & 93.79                                                   & 99.63                                                          & 94.06                                                           & 88.14                                                        & 99.56                                                       & 91.55                                                        & 91.57                                                     & 35.38                                                        & 90.45                                                         & 92.40                                                      \\ \cline{2-15} 
2       & ResNet50V2   & 88.66    & 70.96                                                     & 82.78                                                      & 92.54                                                   & 99.41                                                          & 92.72                                                           & 83.98                                                        & 99.33                                                       & 88.83                                                        & 89.17                                                     & 24.71                                                        & 87.95                                                         & 89.89                                                      \\ \cline{2-15} 
        & Concatenated & 91.56    & 74.19                                                     & 88.52                                                      & 93.60                                                   & 99.76                                                          & 93.69                                                           & 88.95                                                        & 99.69                                                       & 91.67                                                        & 91.77                                                     & 46                                                           & 90.01                                                         & 92.87                                                      \\ \hline
        & Xception     & 91.79    & 70                                                        & 89.18                                                      & 93.57                                                   & 99.75                                                          & 93.66                                                           & 89.6                                                         & 99.67                                                       & 91.91                                                        & 92.01                                                     & 42.85                                                        & 90.04                                                         & 93.26                                                      \\ \cline{2-15} 
3       & ResNet50V2   & 90.47    & 73.33                                                     & 85.29                                                      & 93.89                                                   & 99.14                                                          & 94.30                                                           & 86.81                                                        & 99.07                                                       & 90.78                                                        & 91.11                                                     & 18.48                                                        & 90.58                                                         & 91.63                                                      \\ \cline{2-15} 
        & Concatenated & 91.79    & 83.33                                                     & 87.03                                                      & 94.90                                                   & 99.69                                                          & 95.03                                                           & 87.64                                                        & 99.65                                                       & 91.90                                                        & 92.04                                                     & 41.66                                                        & 91.83                                                         & 92.20                                                      \\ \hline
        & Xception     & 90.70    & 70.96                                                     & 86.38                                                      & 93.57                                                   & 99.63                                                          & 93.71                                                           & 87.06                                                        & 99.55                                                       & 90.84                                                        & 91.01                                                     & 34.37                                                        & 89.81                                                         & 91.75                                                      \\ \cline{2-15} 
4       & ResNet50V2   & 89.38    & 70.96                                                     & 90.83                                                      & 88.52                                                   & 99.31                                                          & 88.99                                                           & 91.82                                                        & 99.23                                                       & 89.71                                                        & 89.82                                                     & 22                                                           & 84.11                                                         & 94.33                                                      \\ \cline{2-15} 
        & Concatenated & 90.47    & 83.87                                                     & 87.33                                                      & 92.54                                                   & 99.32                                                          & 93.03                                                           & 88.34                                                        & 99.27                                                       & 90.8                                                         & 90.89                                                     & 25.24                                                        & 88.94                                                         & 92.43                                                      \\ \hline
        & Xception     & 91.99    & 67.74                                                     & 91.42                                                      & 92.46                                                   & 99.64                                                          & 92.71                                                           & 91.87                                                        & 99.55                                                       & 92.20                                                        & 92.23                                                     & 33.87                                                        & 88.95                                                         & 94.59                                                      \\ \cline{2-15} 
5       & ResNet50V2   & 90.01    & 67.74                                                     & 81.53                                                      & 95.59                                                   & 99.63                                                          & 95.87                                                           & 81.98                                                        & 99.54                                                       & 90.27                                                        & 90.23                                                     & 33.33                                                        & 92.69                                                         & 89.08                                                      \\ \cline{2-15} 
        & Concatenated & 92.08    & 77.41                                                     & 89.16                                                      & 94.03                                                   & 99.62                                                          & 94.33                                                           & 89.62                                                        & 99.56                                                       & 92.31                                                        & 92.29                                                     & 35.82                                                        & 90.99                                                         & 93.30                                                      \\ \hline
        & Xception     & 91.31    & 73.35                                                     & 88.95                                                      & 92.91                                                   & 99.55                                                          & 93.17                                                           & 89.63                                                        & 99.48                                                       & 91.52                                                        & 91.62                                                     & 33.39                                                        & 89.35                                                         & 93.26                                                      \\ \cline{2-15} 
Average & ResNet50V2   & 89.79    & 74.02                                                     & 85.54                                                      & 92.60                                                   & 99.33                                                          & 92.98                                                           & 86.64                                                        & 99.26                                                       & 90.07                                                        & 90.25                                                     & 24.09                                                        & 88.85                                                         & 91.50                                                      \\ \cline{2-15} 
        & Concatenated & 91.40    & 80.53                                                     & 87.35                                                      & 94.06                                                   & 99.56                                                          & 94.32                                                           & 88.09                                                        & 99.50                                                       & 91.60                                                        & 91.71                                                     & 35.27                                                        & 90.83                                                         & 92.40                                                      \\ \hline
\end{tabular}}
\end{table*}

\newpage
\section{Discussion}
\label{5}
It can be understood from the confusion matrices and the tables that the concatenated network performs better in detecting COVID-19 and not detecting false cases of COVID-19 and outputs better overall accuracy.
 {Although we had an unbalanced dataset and a few cases of COVID-19, by using the proposed technique, we could have improved COVID-19 detection along with the other classes detection.}
The reason the precision of COVID-19 class is low is that in our work, despite some other researches that worked on detecting COVID-19 from X-ray images, we tested our neural nets on a massive number of images. Our test images were much more than our train images. As it is explained above,  because we had only 31 cases of COVID-19 and 11271 cases from the other two classes, the false positives of the COVID-19 class would become more than true positives. For example, in the first fold, the concatenated network detected 26 cases correctly out of 31 COVID-19 cases, and from 11271 other cases, only mistakenly identify 68 cases as COVID-19. If we had equal samples from COVID-19 class as the other classes, the precision would become high value. Still, because having few COVID-19 cases and lots of other cases for validation, the precision would become a low value. 

In another study, the results were presented in two forms, 2 and 3 classes, that due to the imbalance in the dataset, there are several meaningless results \cite{apostolopoulos2020covid}.  {We have presented the results for each class and for all the classes with meaningful results that are more practical.} We could have tested our network on a few cases like some of the other researches done recently, but we wanted to show the real performance of our network with few COVID-19 cases. As mentioned, mistakenly detecting 68 cases from 11271 cases to be infected to COVID-19 is not very much but not very well also, and we hope that by using much-provided data from patients infected, COVID-19, the detection accuracy will rise much more.

\section{Conclusion}
\label{6}
In this paper, we presented a concatenated neural network based on Xception and ResNet50V2 networks for classifying the chest X-ray images into three categories of normal, pneumonia, and COVID-19. We used two open-source datasets that contained 180 and 6054 images from patients infected to COVID-19 and pneumonia, respectively, and 8851 images from normal people.
  {As we had a few images of COVID-19 class, we proposed a method for training the neural network when we the dataset is unbalanced. We separated the training set into 8 successive phases, in which there were 633 images(149 COVID-19, 234 pneumonia, 250 normal)  in each phase.}
 {We selected the number of each class almost equal to each other in each phase so that our network also learns COVID-19 class characteristics, not only the other two class features. In each phase, the images from normal and pneumonia classes were different so that the network can distinguish COVID-19 from other classes better.} Our training set included 3783 images, and the rest of the images were allocated for evaluating the network.  We tried to test our model on a large number of images so that our real achieved accuracy be clear. We achieved an average accuracy of 99.50\%, and 80.53\%  {sensitivity} for the COVID-19 class, and the overall accuracy equal to 91.4\% between five folds.  {We hope that our trained network that is publicly available be helpful for medical diagnosis. We also hope that in the future, larger datasets from COVID-19 patients become available, and by using them, the accuracy of our proposed network increases for good.}

\section{Code Availability}
\label{7}
In this GitHub profile (\url{https://github.com/mr7495/covid19}), we have shared the trained networks and all the used code in this paper. We hope our work be useful to help in future researches.

\section*{Acknowledgment}
We would like to appreciate Joseph Paul Cohen and the others who provided these x-ray images from patients infected to COVID-19.
We thank Linda Wang and Alexander Wong for making their code available, in which we have used a part of it on our research for preparing our dataset. We also thank \href{https://colab.research.google.com/}{Google Colab server} for providing free GPU.

This is a preprint of an article published in Informatics in Medicine Unlocked journal. The final authenticated version is available online at \href{https://doi.org/10.1016/j.imu.2020.100360}{10.1016/j.imu.2020.100360}.

\bibliographystyle{abbrv}
\bibliography{arxiv}

\end{document}